\documentclass[prl,aps,floats,twocolumn,prd,showpacs,nofootinbib]{revtex4}

\usepackage{amssymb}
\usepackage{amsmath}

\usepackage{graphicx}
\usepackage{graphics}
\usepackage{dcolumn}
\usepackage{color}
\usepackage{rotate}
\usepackage{fancyhdr}

\begin{document}

\title{The Pesky Power Asymmetry}

\author{Liang Dai, Donghui Jeong, Marc Kamionkowski and Jens Chluba}
\affiliation{Department of Physics and Astronomy, Johns
     Hopkins University, 3400 N.\ Charles St., Baltimore, MD 21218}

\date{March 27, 2013}

\begin{abstract}
 Physical models for the hemispherical power asymmetry in the
 cosmic microwave background (CMB) reported by the Planck
 Collaboration must satisfy CMB constraints to the 
 homogeneity of the Universe and quasar constraints to
 power asymmetries.  We survey a variety of models for the power
 asymmetry and show that consistent models include a
 modulated scale-dependent isocurvature contribution to the
 matter power spectrum or a modulation of the reionization
 optical depth, gravitational-wave amplitude, or scalar spectral
 index. We propose further tests to distinguish between the
 different scenarios.
\end{abstract}


\maketitle

The Planck Collaboration has reported a hemispherical asymmetry
in the cosmic microwave background (CMB) fluctuations \cite{Ade:2013sta}, thus 
confirming a similar power asymmetry seen in the Wilkinson
Microwave Anisotropy Probe (WMAP) data
\cite{Eriksen:2003db}.  The new data, with
far better multifrequency component separation, make it more
difficult to attribute the asymmetry to foregrounds.  The
asymmetry is also seen in the Planck data to extend to smaller
scales, and it is thus of greater statistical significance than
in the WMAP data.

If the asymmetry is modeled as a dipolar modulation of
an otherwise statistically isotropic CMB sky
\cite{Prunet:2004zy,Gordon:2005ai,Gordon:2006ag}, the best-fit
dipole has direction $(l,b)=(227,-27)$ and amplitude (in terms
of r.m.s.\ temperature fluctuations on large angular
scales, multipoles $\ell<64$) of $A=0.072\pm0.022$, although the
asymmetry extends to higher $\ell$
\cite{Axelsson:2013mva,Ade:2013sta}.

This power asymmetry is, as we explain below, extremely
difficult to reconcile with inflation.  Given the plenitude of
impressive successes of inflation (the nearly, but not
precisely, Peebles-Harrison-Zeldovich spectrum, adiabatic rather
than isocurvature perturbations, the remarkable degree of
Gaussianity), the result requires the deepest scrutiny.  While
there are some who may wave away the asymmetry as a 
statistical  fluctuation \cite{Bennett:2010jb}, evidence is
accruing that it is statistically significant.  There is
moreover the tantalizing prospect that it may be an artifact of
some superhorizon pre-inflationary physics.  Here we investigate
physical explanations for the origin of the asymmetry and
put forward new tests of those models.

We begin by reviewing the tension between the
asymmetry and inflation.  We then provide a brief survey of
prior hypotheses and discuss the very stringent constraints
imposed by the CMB temperature quadrupole and by upper limits to
hemispherical asymmetries in quasar abundances.  We review some
existing models for the asymmetry and then posit that
the asymmetry may be due to spatial variation of 
standard cosmological parameters
(e.g., the reionization optical depth, the scalar spectral
index, and gravitational-wave amplitude) that affect CMB
fluctuations without affecting the total density nor the matter
power spectrum.  We show how the different scenarios can be
distinguished by the $\ell$ dependence of the asymmetry, and we
discuss other possible tests of the models.

The CMB power asymmetry is modeled as a dipole modulation of the
power; i.e., the temperature fluctuation in direction
$\hat n$ is $(\Delta T/T)(\hat n) = s(\hat n) [1 + A \, \hat n
\cdot \hat p]$, where $s(\hat n)$ is a statistically isotropic
map, $A$ is the power-dipole amplitude, and
$\hat p$ is its direction.  However, the asymmetry cannot arise
due to a preferred direction in the three-dimensional spectrum $P(k)$
\cite{Pullen:2007tu}, as reality of the fractional
matter-density perturbation $\delta({\bf r})$ relates the Fourier
component $\tilde \delta({\bf k})$ for wavevector ${\bf k}$ to
that, $\tilde\delta(-{\bf k})=\tilde\delta^*({\bf k})$, of $-{\bf k}$.  The
asymmetry must therefore be attributed to a {\it spatial
modulation} of three-dimensional power across the observable
Universe.  The modulation required to explain the asymmetry can
then be written in terms of a spatially-varying power spectrum,
$P(k,{\bf r})=P(k) [1 + 2 A \, \hat p \cdot {\bf r}/r_{ls}]$,
where $r_{ls}$ is the distance to the last-scattering surface,
for modes inside the comoving horizon at present ($k\gtrsim H_0^{-1}$).

Any model that modulates the power must do so without
introducing a modulation in the density of the Universe.
A long-wavelength isocurvature density fluctuation with an
amplitude $O(10\%)$ must generates a temperature dipole two
orders of magnitude greater than is observed.  If the density
fluctuation is adiabatic, then the intrinsic temperature dipole
is cancelled by a Doppler dipole due to our infall toward the
denser side.  Even so, in this case, the small temperature
quadrupole and octupoles constrain the density in the observed
Universe to vary by no more than $O(10^{-3})$
\cite{Grishchuk:1978}.

These considerations make it unlikely that the power asymmetry
could arise in single-clock models for inflation.  In these
models, the inflaton controls both the power-spectrum amplitude
and the total density, thus making it difficult to introduce an
$O(10\%)$ modulation in the power with a $\lesssim O(10^{-3})$
modulation of the total density.  These arguments were made
precise for slow-roll inflation with a standard kinetic term in
Ref.~\cite{Erickcek:2008sm}.  We surmise that it may be
difficult to get to work also with nontrivial kinetic terms
\cite{Silverstein:2003hf}, especially given
the increasingly tight constraints to such models from 
Planck.

There are additional and very stringent constraints to power
modulation---even in models that can do so without modulating the
density---from the Sloan Digital Sky Survey (SDSS) quasar sample
\cite{Hirata:2009ar}.  These quasars are found at distances
nearly half of the comoving distance to the CMB surface of last
scatter, and their abundances depend very sensitively on the
amplitude of the primordial power spectrum.  A detailed analysis
\cite{Hirata:2009ar}
finds an upper limit $A<0.0153$ (99\% C.L.) to the amplitude of
an asymmetry oriented in the direction of the CMB dipole.  While
this is consistent with the $3\sigma$ lower limit
$A\gtrsim0.006$ inferred from Ref.~\cite{Hoftuft:2009rq}, it is
roughly five times smaller than the central value $A=0.072$.  Of
course, quasars probe the power 
spectrum primarily at wavenumbers $k\sim1$~Mpc$^{-1}$, while the
$\ell\lesssim 60$ CMB power probes $k\lesssim 0.035$~Mpc$^{-1}$.  A
model that produces a power asymmetry with a sufficiently strong
scale dependence may (neglecting the possible extension of the
CMB asymmetry to higher $\ell$) allow the central CMB value to be
consistent with the quasar constraint.

To summarize:  Any mechansim that accounts for the CMB power
asymmetry at $\ell\lesssim60$ must (1) do so while leaving the
amplitude of the long-wavelength adiabatic or isocurvature density fluctuation
across the observable Universe to be $\lesssim 10^{-3}$, 
and (2) leave the power asymmetry at $k\sim1$~Mpc$^{-1}$ small.  Given that the
asymmetries are seen at $\ell<60$, it is also likely that any
causal mechanism must involve inflation.  We now run through a
number of scenarios for the power asymmetry.

Ref.~\cite{Erickcek:2008sm} arranged for a scale-independent power
modulation, while keeping the total density fixed, by
introducing a curvaton during inflation.  The curvaton can
then contribute appreciably to a modulation of the
density-perturbation amplitude without modulating the mean density.
This model is inconsistent with the quasar bound for the
best-fit value $A=0.072$.  This model also predicts a
non-Gaussianity parameter 
${f_{\rm NL}^{\mathrm{(local)}}} \gtrsim 25 (A/0.07)^2$,
and would thus be ruled out for $A=0.072$ by Planck constraints
\cite{Ade:2013ydc} to ${f_{\rm NL}^{(\mathrm{local})}}$, even if there were no quasar
constraint.

Ref.~\cite{Erickcek:2009at} also presented a modified inflationary
theory wherein the curvaton decays after dark matter freezes out
thus giving rise to an isocurvature perturbation that is
subdominant relative to the usual adiabatic perturbations from
inflaton decay.  A postulated superhorizon perturbation to the
curvaton field then modulates the amplitude of the isocurvature
contribution across the observable Universe.  The model
parameters can be chosen so that this spatially-varying
isocurvature contribution is scale-dependent, with a CMB power
spectrum that peaks around $\ell\sim10$, falls off rapidly from
$\ell\sim10$ to $\ell\sim 100$, and is then negligibible at higher
$\ell$.  The model predicts an isocurvature contribution to
primordial perturbations that may be in tension with new Planck
upper limits \cite{Ade:2013rta}, although more analysis
may be required, given the asymmetric nature of the
contribution, to determine the consistency of the model with
current Planck data.

Ref.~\cite{Schmidt:2012ky} recently postulated that the power
asymmetry may arise in single-field inflation through some sort
of non-Gaussianity that increases the bispectrum
in the squeezed limit.  In this way, the homogeneity constraint
imposed by the CMB can be evaded.  This model, however, gives
rise to a roughly scale-independent power asymmetry and thus
conflicts with the quasar constraint.  It may still be
possible though, to adjust the parameters to reduce the
asymmetry on small scales.

\begin{figure}[htbp]
\includegraphics[width=3.7in]{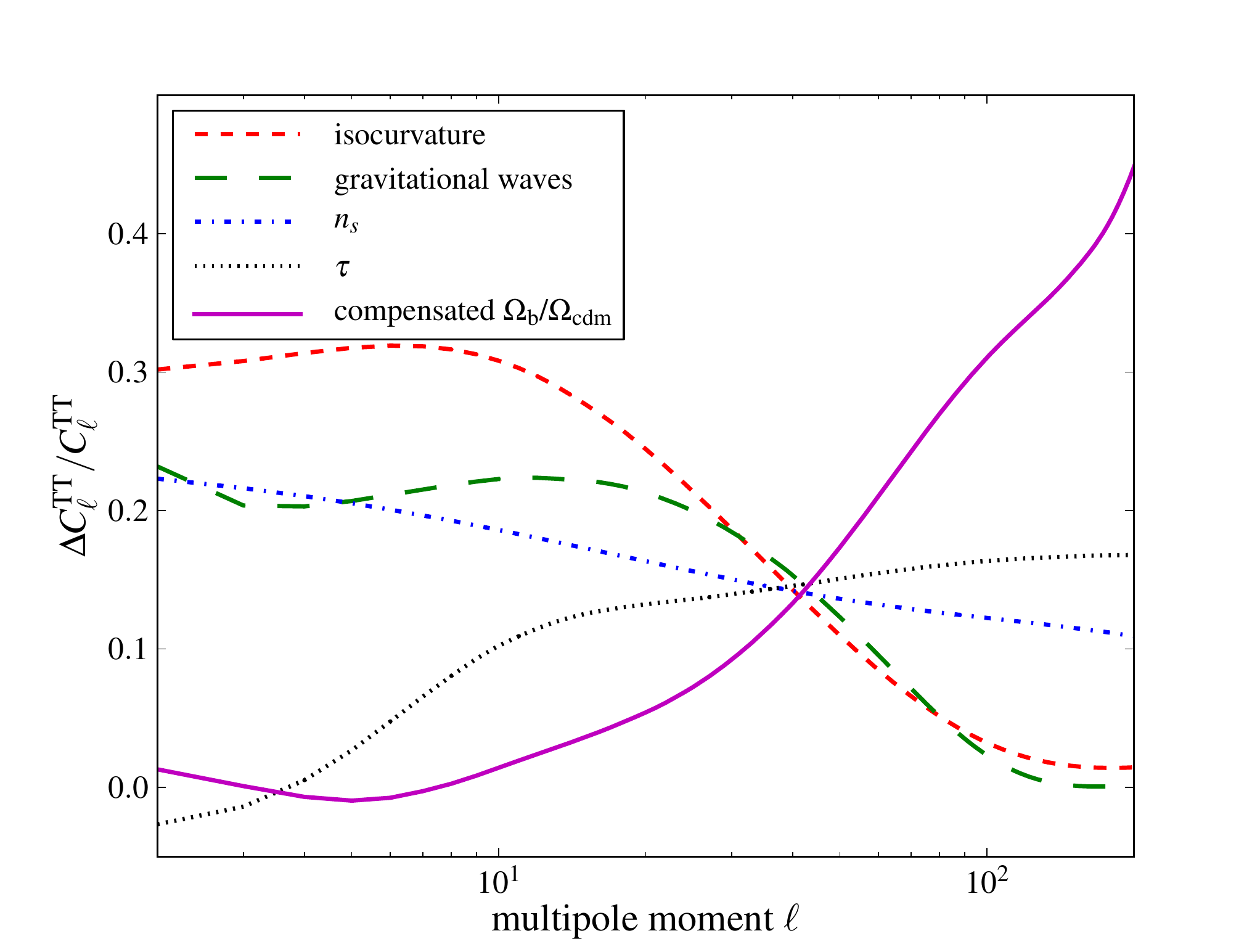}
\caption{The fractional change $\Delta C_{\ell}^{\rm TT}/C_{\ell}^{\rm
     TT}$ in the CMB power spectrum due to the inflationary
     model of Ref.~\protect\cite{Erickcek:2009at}
     and modulation of the gravitational-wave amplitude, scalar
     spectral index, reionization optical depth, and baryon
     density (compenstated by the dark-matter density so that
     the total density is fixed).  Each curve is normalized so
    that $A=0.072$.}
\label{fig:DeltaCls}
\end{figure}
\bigskip

\begin{figure}[htbp]
\includegraphics[width=3.7in]{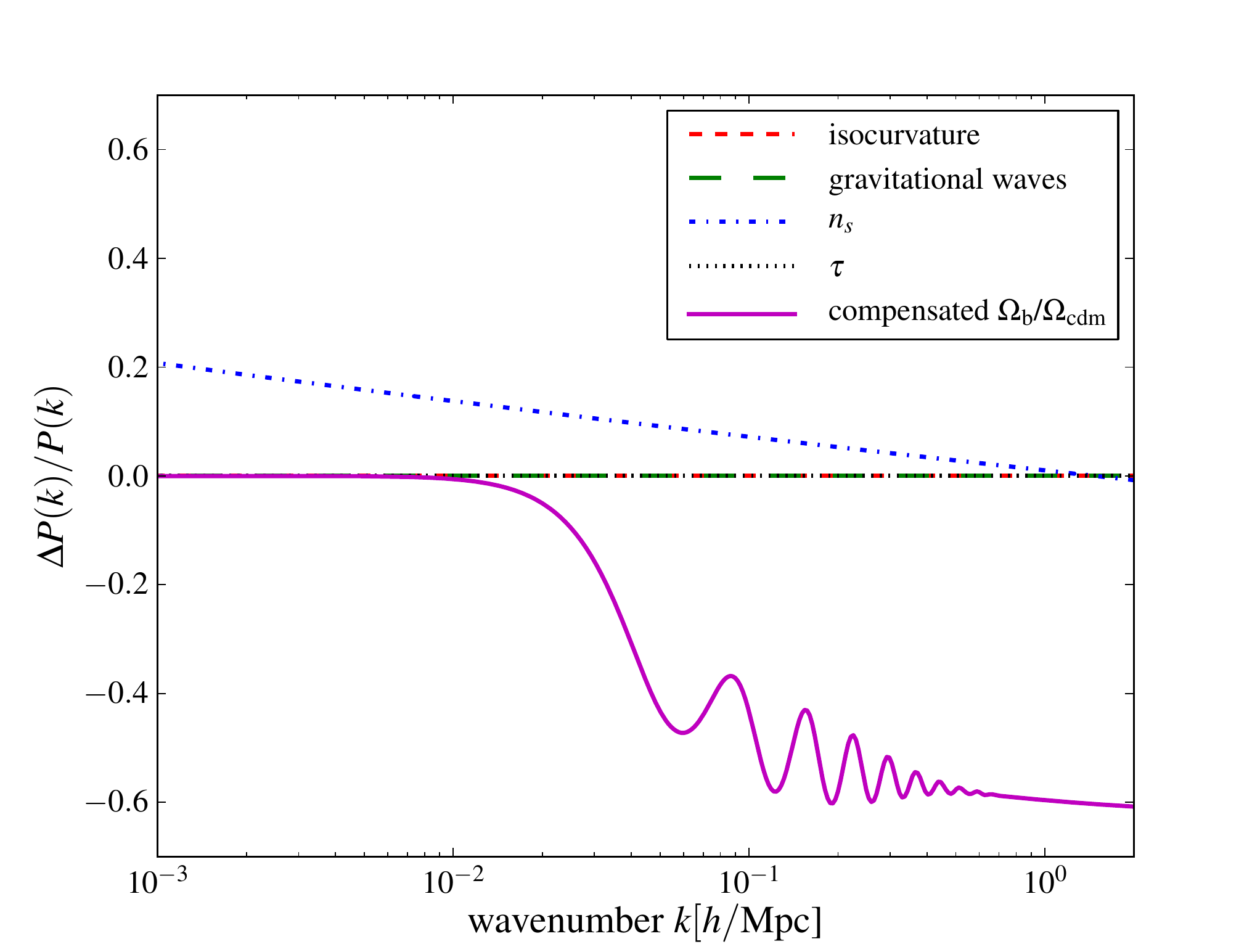}
\caption{The fractional change $\Delta P(k)/P(k)$ in the matter
     power spectrum for each of the models shown in
     Fig.~\protect\ref{fig:DeltaCls}.}
\label{fig:DeltaPk}
\end{figure}
\bigskip

We now suppose that the power asymmetry may
arise from a modulation of one of the cosmological parameters
that affects the CMB power spectrum.  There
are a number of cosmological parameters---abundances of cosmic constituents,
inflationary observables, fundamental-physics parameters
\cite{Moss:2010qa}---that affect the CMB power spectrum
\cite{Jungman:1995bz}.  If there is a difference between the
value of one of these cosmological parameters on one side of the
sky and the value on the other side, then the CMB power spectrum
one side may differ from that on the other side.
For each of these parameters $p$, we calculate $\Delta
C_\ell^{\rm TT}=(\partial C_\ell^{\rm TT}/\partial p) \Delta p$, where the amplitude
$\Delta p$ is chosen so that it produces an asymmetry
$A=0.072$. We assume here that this asymmetry is determined from
the data by weighting the asymmetry in all spherical-harmonic
modes equally up to $\ell_{\rm max}=64$; i.e., 
\begin{equation}
     A = \frac{ (1/2) \sum_{\ell=2}^{\ell_{\rm max}} (2\ell+1) (\Delta
     C_\ell^{\rm TT}/C_\ell^{\rm TT}) }{ \sum_{\ell=2}^{\ell_{\rm max}} (2\ell+1)}.
\end{equation}
The fractional power-spectrum differences, normalized to give
$A=0.072$ are given in Fig.~\ref{fig:DeltaCls}.
We then plot in Fig.~\ref{fig:DeltaPk} the fractional change $\Delta P(k)$
in the matter power spectrum induced by the modulations of each
of the parameters considered in Fig.~\ref{fig:DeltaPk}.  We
do not consider parameters that only affect the recombination
history (e.g., the helium abundance, fine-structure constant,
etc.), as these modify CMB power only on small scales (high
$\ell$), not at the lower $\ell$ where the asymmetry is best seen.

Ref.~\cite{Axelsson:2013mva} considers, among other
possibilities, a modulation in $\Omega_{\rm b}$, the baryon density.
However, if $\Omega_{\rm b}$ is modulated by $O(10\%)$, while holding
all other parameters fixed, it will introduce a
large-scale inhomogeneity in conflict with the CMB
dipole/quadrupole/octupole.  These constraints can be evaded
through a compensated isocurvature perturbations (CIP), wherein
$\Omega_{\rm b}$ and $\Omega_{\rm cdm}$, the dark-matter density, are
both modulated in such a way that the total matter density
$\Omega_{\rm b}+\Omega_{\rm cdm}$ remains constant across the
observable Universe \cite{CIP}.  Such a hypothesis results, as
Fig.~\ref{fig:DeltaPk} shows, in a power asymmetry at small
scales larger than allowed by the quasar constraint.  It can
therefore be ruled out.

Ref.~\cite{Gordon:2005ai} considered a dark-energy
density that varies linearly with
position along the direction picked out by the power dipole.
The dark-energy density is negligible at the surface of last
scatter, but it affects at later times CMB fluctuations in two
different ways.  First of all, changes to $\Omega_\mathrm{de}$ 
change the ISW contribution
to low-$\ell$ power, but this is a relatively small effect.  The
other consequence is a change to the angle subtended by a given
comoving scale.  The effect of a dipolar modulation, across the
sky, of this mapping is equivalent to, and indistinguishable
from, that induced by a peculiar velocity.  A variation of 
$\Omega_{\mathrm{de}}$
large enough to account for the $A\simeq0.072$ asymmetry would
thus yield a CMB temperature dipole considerably larger than
that observed.  This explanation can thus be learned out.

More generally, a modulation of any of the parameters that
affects the total density of the Universe that is large enough
to account for the power asymmetry will give rise to a
large-angle CMB fluctuation in gross conflict with observations.
We thus now consider modulations to several parameters that
affect the CMB power spectrum without changing the matter
densities.

We begin with a variation to the scalar spectral index $n_{\rm
s}$.  In considering a modulation of $n_s$, we must specify a
pivot wavenumber $k_0$, at which the power on both sides 
of the sky is equal.  Here we choose this pivot point to be
$k_0=1$~Mpc$^{-1}$ so that the quasar constraint is satisfied.
Doing so allows us accommodate a large-scale power-asymmetry
amplitude $A=0.072$ with a value of $n_{\rm s}\simeq 0.93$ on one side
of the sky and $n_{\rm s}\simeq 0.99$ on the other.  This model then
predicts that the CMB power asymmetry should decrease, but
relatively slowly, with higher $\ell$, as shown in Fig.~\ref{fig:DeltaCls}.

Along similar lines, one can vary the gravitational-wave
amplitude from one side of the sky to the other.  The
gravitational-wave energy-density flucutation required to
account for the low-$\ell$ power asymmetry is small enough to
satisfy the homogeneity constraints, and gravitational waves
contribute nothing to $P(k)$.  This hypothesis can thus explain
the CMB power asymmetry without violating other constraints.
The only difficulty with the model is that an asymmetry
amplitude $A=0.072$ requires a gravitational-wave amplitude on
one side of the Universe ten times larger than the homogeneous
upper limit, a magnitude that may be not only unpalatable, but
also inconsistent with current data.  Still, an asymmetry of
smaller amplitude, say $A\sim0.015$ may be consistent.

We finally consider a dipolar modulation of $\tau$, the
reionization optical depth.  The optical depth primarily
suppresses power at $\ell\gtrsim20$, but there is also a small
increase in power at lower $\ell$ from re-scattering
of CMB photons.  We find from Figs.~\ref{fig:DeltaCls} and
\ref{fig:DeltaPk} that a modulation of $\tau$
can account for the asymmetry in the CMB without affecting (by
assumption, really) $P(k)$ at quasar scales.  An asymmetry
$A=0.072$ can be obtained by taking $\tau=0.017$ on one side of
the Universe and $\tau=0.21$ on the other side. 
 A value of $\tau\simeq0.017$ implies a reionization redshift $z \simeq3$,
which is lower than quasar absorption spectra allow.
Still, an asymmetry $A\simeq0.05$ could be accommodated while preserving 
a reionization redshift $z\gtrsim6$ everywhere.
We surmise, without developing a detailed  microphysical model that a spatial
modulation in the primordial $P(k)$ at $k\gg$Mpc$^{-1}$ could
give rise to such a $\tau$ modulation, or perhaps one of
slightly lower amplitude, given the highly
uncertain physics reionization and the possibly strong
dependence to small changes in initial conditions.

To summarize, we have four models that can account for the CMB
asymmetry while leaving the Universe homogeneous and without
affecting $P(k)$ on quasar scales.  There is the inflation model
of Ref.~\cite{Erickcek:2009at} whose possible phenomenological
weakness may be in an isocurvature contribution in tension with
current upper limits, and there is modulation of $n_s$, which is
phenomenologically quite attractive.  The other two
models---variable gravitational-wave amplitude and variable
$\tau$---require variations in the parameters that are probably
larger than are allowed.  Still, we continue to consider them
for illustrative purposes and in case the asymmetry amplitude is
found in the future to be smaller but still nonzero.
We now discuss measurements that can distinguish between the
different scenarios.

\begin{figure}[htbp]
\includegraphics[width=3.7in]{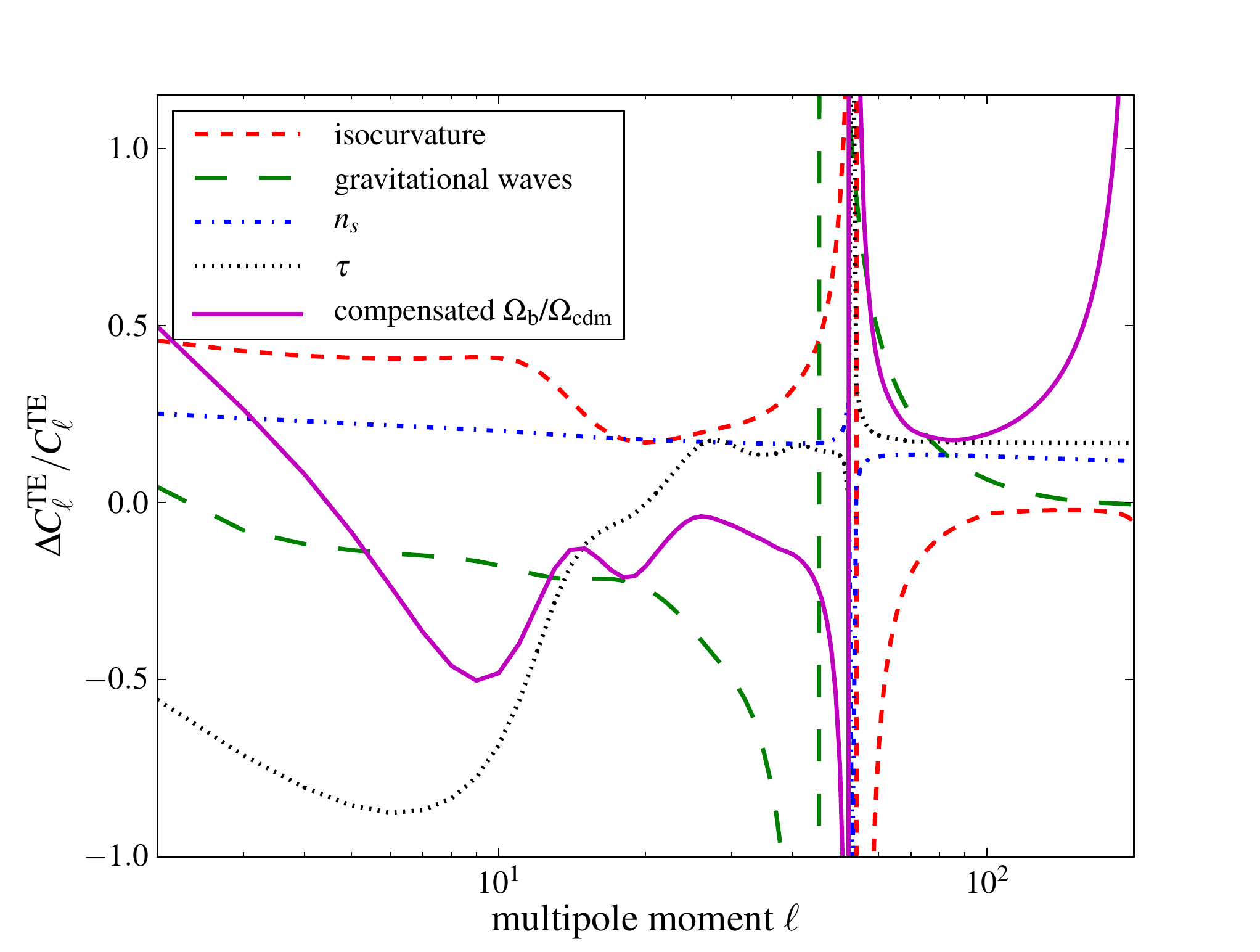}
\caption{The fractional changes $\Delta C_\ell^{\rm TE}/C_\ell^{\rm
     TE}$ in the CMB temperature-polarization power spectrum for
     each of the models shown in Fig.~\protect\ref{fig:DeltaCls}.}
\label{fig:DeltaTE}
\end{figure}
\bigskip

\begin{figure}[htbp]
\includegraphics[width=3.7in]{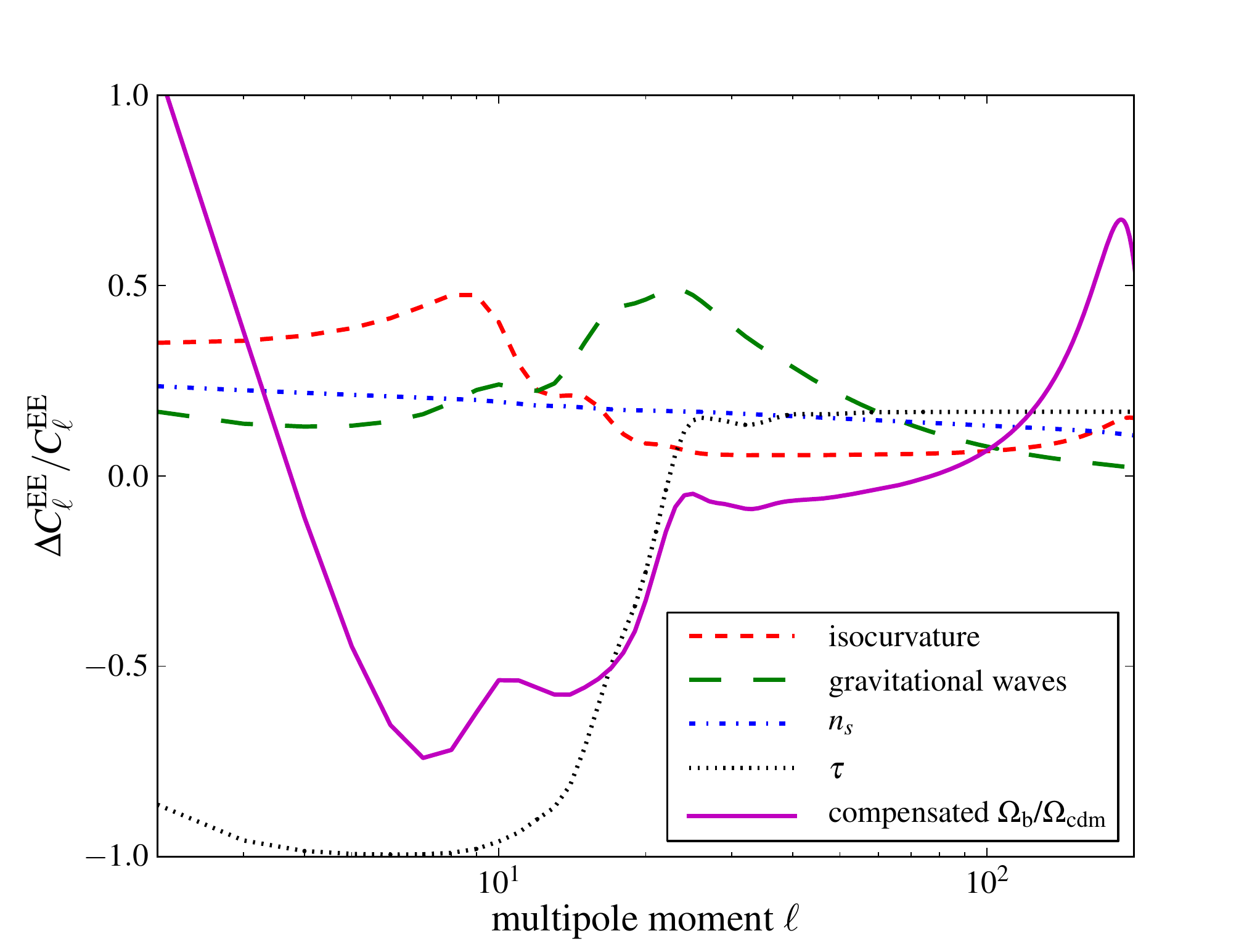}
\caption{The fractional changes $\Delta C_\ell^{\rm EE}/C_\ell^{\rm
     EE}$ in the CMB polarization power spectrum for
     each of the models shown in
     Fig.~\protect\ref{fig:DeltaCls}.}
\label{fig:DeltaEE}
\end{figure}
\bigskip

First of all, with Plank data we should be able to measure the
difference $\Delta C_\ell^{\rm TT}$ in the power spectra between the two
hemispheres, as a function of $\ell$.  The model of
Ref.~\cite{Erickcek:2009at} and a
modulation of the gravitational-wave amplitude both predict
little or no asymmetry at $\ell\gtrsim100$, while the power
asymmetry should extend to much higher $\ell$ if it is due to a
modulation of $n_{\rm s}$ or $\tau$.  There are also the
temperature-polarization correlations ($C_\ell^{\rm TE}$) and
polarization autocorrelations ($C_\ell^{\rm EE}$) shown in
Figs.~\ref{fig:DeltaTE} and \ref{fig:DeltaEE}.  The TE
difference power spectrum, in particular, should help
distinguish, through the sign at low $\ell$ of $\Delta C_\ell^{\rm TE}$,
modulation of gravitational waves from modulation of
$\tau$.  An asymmetry in $P(k)$ can be probed at
lower $k$ than quasars probe through all-sky lensing,
Compton-$y$, and/or cosmic-infrared-background (CIB) maps, such
as those recently made by Planck
\cite{Ade:2013nta}.
Asymmetries at $\sim0.1-1$~Mpc$^{-1}$ scales may also be probed
via number counts in populations of objects other than quasars
\cite{Gibelyou:2012ri}.
Probing asymmetries in $P(k)$ at $k\gg$Mpc$^{-1}$, as required
if $\tau$ is modulated, may be more
difficult in the near term, although future 21-cm maps of the
neutral-hydrogen distribution during the dark ages
\cite{Loeb:2003ya} and/or epoch of reionization
\cite{21cm} may do the trick, as may maps of the $\mu$
distortion to the CMB frequency spectrum \cite{Chluba:2012we}.
A modulation in the gravitational-wave background may show up
as an asymmetry in CMB B modes \cite{Kamionkowski:1996zd}.  In
fact, if the asymmetry is attributed to a gravitational-wave
asymmetry, suborbital B-mode searches may do better to search on
one side of the sky than on the other!  If the asymmetry has an
origin in the coupling of an inflaton to some other field, a
``fossil'' field, during inflation, there may be higher-order
correlation functions at smaller scales that can be sought
\cite{Jeong:2012df}.  It may also be instructive to perform a
 bipolar spherical harmonic (BiPoSH) \cite{Hajian:2004zn} analysis
and consider the odd-parity dipolar (i.e., $L=1$, with
$\ell+\ell'+L$=odd) BiPoSH \cite{Book:2011na}, which may shed light on
the nature of a fossil field \cite{Dai:2013ikl} that would give
rise to the asymmetry.  

While a modulation in $n_{\rm s}$ can account for the CMB power
asymmetry, and a modulation to the gravitational-wave amplitude
may do so for a smaller-amplitude asymmetry, these are both no 
more than phenomenological hypotheses, and there may be
difficult work ahead to accommodate them within an inflationary
model.  Similar comments apply to variable optical depth.

There may also well be a completely
different explanation, like bubble collisions
\cite{Chang:2008gj} or non-trivial topology of the Universe
\cite{Ade:2013vbw}, that is not well described by the
modulation models we have considered here.  Any such model must
still, however, satisfy the CMB constraints to homogeneity and
the quasar limit to a small-scale power asymmetry.
Finally, if the asymmetry does indeed signal
something beyond the simplest inflationary models, then it may
be possible to draw connections between it and the tension
between CMB and local models of the Hubble constant, between
different suborbital CMB experiments, and perhaps other anomalies in
current data.

\smallskip

This work was supported by DoE SC-0008108 and NASA NNX12AE86G.


\begin{thebibliography}{99}

%\cite{Ade:2013sta}
\bibitem{Ade:2013sta} 
  P.~A.~R.~Ade {\it et al.}  (Planck Collaboration),
  %``Planck 2013 results. XXIII. Isotropy and Statistics of the CMB,''
  arXiv:1303.5083 [astro-ph.CO].
  %%CITATION = ARXIV:1303.5083;%%

%\cite{Eriksen:2003db}
\bibitem{Eriksen:2003db} 
  H.~K.~Eriksen {\it et al.}, 
  %``Asymmetries in the Cosmic Microwave Background anisotropy field,''
  Astrophys.\ J.\  {\bf 605}, 14 (2004)
  [Erratum-ibid.\  {\bf 609}, 1198 (2004)]
  [astro-ph/0307507];
  %%CITATION = ASTRO-PH/0307507;%%
  %346 citations counted in INSPIRE as of 22 Mar 2013
  H.~K.~Eriksen {\it et al.}
  %``Hemispherical power asymmetry in the three-year Wilkinson Microwave Anisotropy Probe sky maps,''
  Astrophys.\ J.\  {\bf 660}, L81 (2007)
  [astro-ph/0701089].
  %%CITATION = ASTRO-PH/0701089;%%
  %109 citations counted in INSPIRE as of 22 Mar 2013

%\cite{Prunet:2004zy}
\bibitem{Prunet:2004zy} 
  S.~Prunet, J.~-P.~Uzan, F.~Bernardeau and T.~Brunier,
  %``Constraints on mode couplings and modulation of the CMB with WMAP data,''
  Phys.\ Rev.\ D {\bf 71}, 083508 (2005)
  [astro-ph/0406364].
  %%CITATION = ASTRO-PH/0406364;%%
  %49 citations counted in INSPIRE as of 20 Mar 2013

%\cite{Gordon:2005ai}
\bibitem{Gordon:2005ai} 
  C.~Gordon, W.~Hu, D.~Huterer and T.~M.~Crawford,
  %``Spontaneous isotropy breaking: a mechanism for cmb multipole alignments,''
  Phys.\ Rev.\ D {\bf 72}, 103002 (2005)
  [astro-ph/0509301].
  %%CITATION = ASTRO-PH/0509301;%%
  %90 citations counted in INSPIRE as of 20 Mar 2013

%\cite{Gordon:2006ag}
\bibitem{Gordon:2006ag} 
  C.~Gordon,
  %``Broken Isotropy from a Linear Modulation of the Primordial Perturbations,''
  Astrophys.\ J.\  {\bf 656}, 636 (2007)
  [astro-ph/0607423].
  %%CITATION = ASTRO-PH/0607423;%%
  %20 citations counted in INSPIRE as of 20 Mar 2013

%\cite{Axelsson:2013mva}
\bibitem{Axelsson:2013mva} 
  M.~Axelsson {\it et al.},
  %``Directional dependence of LCDM cosmological parameters,''
  arXiv:1303.5371 [astro-ph.CO].
  %%CITATION = ARXIV:1303.5371;%%

%\cite{Bennett:2010jb}
\bibitem{Bennett:2010jb} 
  C.~L.~Bennett, {\it et al.} (WMAP Collaboration),
  %``Seven-Year Wilkinson Microwave Anisotropy Probe (WMAP) Observations: Are There Cosmic Microwave Background Anomalies?,''
  Astrophys.\ J.\ Suppl.\  {\bf 192}, 17 (2011)
  [arXiv:1001.4758 [astro-ph.CO]].
  %%CITATION = ARXIV:1001.4758;%%
  %187 citations counted in INSPIRE as of 22 Mar 2013


%\cite{Pullen:2007tu}
\bibitem{Pullen:2007tu} 
  A.~R.~Pullen and M.~Kamionkowski,
  %``Cosmic Microwave Background Statistics for a Direction-Dependent Primordial Power Spectrum,''
  Phys.\ Rev.\ D {\bf 76}, 103529 (2007)
  [arXiv:0709.1144 [astro-ph]].
  %%CITATION = ARXIV:0709.1144;%%
  %107 citations counted in INSPIRE as of 19 Mar 2013

\bibitem{Grishchuk:1978}
  L.~P.~Grishchuk and Ya.~B.~Zel'dovich, Astron.\ Zh.\ {\bf 55},
  209 (1978) [Sov.\ Astron.\ {\bf 22}, 125 (1978)];
  M.~S.~Turner,
  %``A Tilted universe (and other remnants of the preinflationary universe),''
  Phys.\ Rev.\ D {\bf 44}, 3737 (1991);
  %%CITATION = PHRVA,D44,3737;%%
  %60 citations counted in INSPIRE as of 20 Mar 2013
  A.~L.~Erickcek, S.~M.~Carroll and M.~Kamionkowski,
  %``Superhorizon Perturbations and the Cosmic Microwave Background,''
  Phys.\ Rev.\ D {\bf 78}, 083012 (2008)
  [arXiv:0808.1570 [astro-ph]];
  %%CITATION = ARXIV:0808.1570;%%
  %24 citations counted in INSPIRE as of 19 Mar 2013
  J.~P.~Zibin and D.~Scott,
  %``Gauging the cosmic microwave background,''
  Phys.\ Rev.\ D {\bf 78}, 123529 (2008)
  [arXiv:0808.2047 [astro-ph]].
  %%CITATION = ARXIV:0808.2047;%%
  %14 citations counted in INSPIRE as of 22 Mar 2013

%\cite{Erickcek:2008sm}
\bibitem{Erickcek:2008sm} 
  A.~L.~Erickcek, M.~Kamionkowski and S.~M.~Carroll,
  %``A Hemispherical Power Asymmetry from Inflation,''
  Phys.\ Rev.\ D {\bf 78}, 123520 (2008)
  [arXiv:0806.0377 [astro-ph]].
  %%CITATION = ARXIV:0806.0377;%%
  %61 citations counted in INSPIRE as of 19 Mar 2013

%\cite{Silverstein:2003hf}
\bibitem{Silverstein:2003hf} 
  E.~Silverstein and D.~Tong,
  %``Scalar speed limits and cosmology: Acceleration from D-cceleration,''
  Phys.\ Rev.\ D {\bf 70}, 103505 (2004)
  [hep-th/0310221];
  %%CITATION = HEP-TH/0310221;%%
  %433 citations counted in INSPIRE as of 22 Mar 2013
  M.~Alishahiha, E.~Silverstein and D.~Tong,
  %``DBI in the sky,''
  Phys.\ Rev.\ D {\bf 70}, 123505 (2004)
  [hep-th/0404084].
  %%CITATION = HEP-TH/0404084;%%
  %576 citations counted in INSPIRE as of 22 Mar 2013

%\cite{Hirata:2009ar}
\bibitem{Hirata:2009ar} 
  C.~M.~Hirata,
  %``Constraints on cosmic hemispherical power anomalies from quasars,''
  JCAP {\bf 0909}, 011 (2009)
  [arXiv:0907.0703 [astro-ph.CO]].
  %%CITATION = ARXIV:0907.0703;%%
  %14 citations counted in INSPIRE as of 22 Mar 2013

%\cite{Hoftuft:2009rq}
\bibitem{Hoftuft:2009rq} 
  J.~Hoftuf {\it et al.},
  %``Increasing evidence for hemispherical power asymmetry in the five-year WMAP data,''
  Astrophys.\ J.\  {\bf 699}, 985 (2009)
  [arXiv:0903.1229 [astro-ph.CO]].
  %%CITATION = ARXIV:0903.1229;%%
  %86 citations counted in INSPIRE as of 22 Mar 2013

%\cite{Ade:2013ydc}
\bibitem{Ade:2013ydc} 
  P.~A.~R.~Ade {\it et al.}  (Planck Collaboration),
  %``Planck 2013 Results. XXIV. Constraints on primordial non-Gaussianity,''
  arXiv:1303.5084 [astro-ph.CO].
  %%CITATION = ARXIV:1303.5084;%%

%\cite{Erickcek:2009at}
\bibitem{Erickcek:2009at} 
  A.~L.~Erickcek, C.~M.~Hirata and M.~Kamionkowski,
  %``A Scale-Dependent Power Asymmetry from Isocurvature Perturbations,''
  Phys.\ Rev.\ D {\bf 80}, 083507 (2009)
  [arXiv:0907.0705 [astro-ph.CO]].
  %%CITATION = ARXIV:0907.0705;%%
  %25 citations counted in INSPIRE as of 19 Mar 2013

%\cite{Ade:2013rta}
\bibitem{Ade:2013rta} 
  P.~A.~R.~Ade {\it et al.}  (Planck Collaboration),
  %``Planck 2013 results. XXII. Constraints on inflation,''
  arXiv:1303.5082 [astro-ph.CO].
  %%CITATION = ARXIV:1303.5082;%%

%\cite{Schmidt:2012ky}
\bibitem{Schmidt:2012ky} 
  F.~Schmidt and L.~Hui,
  %``CMB Power Asymmetry from Non-Gaussian Modulation,''
  Phys.\  Rev.\  Lett.\  110, {\bf 011301} (2013)
  [arXiv:1210.2965 [astro-ph.CO]].
  %%CITATION = ARXIV:1210.2965;%%
  %3 citations counted in INSPIRE as of 20 Mar 2013

%\cite{Moss:2010qa}
\bibitem{Moss:2010qa} 
  A.~Moss, D.~Scott, J.~P.~Zibin and R.~Battye,
  %``Tilted Physics: A Cosmologically Dipole-Modulated Sky,''
  Phys.\ Rev.\ D {\bf 84}, 023014 (2011)
  [arXiv:1011.2990 [astro-ph.CO]].
  %%CITATION = ARXIV:1011.2990;%%
  %3 citations counted in INSPIRE as of 20 Mar 2013

%\cite{Jungman:1995bz}
\bibitem{Jungman:1995bz} 
  G.~Jungman, M.~Kamionkowski, A.~Kosowsky and D.~N.~Spergel,
  %``Cosmological parameter determination with microwave background maps,''
  Phys.\ Rev.\ D {\bf 54}, 1332 (1996)
  [astro-ph/9512139].
  %%CITATION = ASTRO-PH/9512139;%%
  %366 citations counted in INSPIRE as of 25 Mar 2013

%\cite{Holder:2009gd}
\bibitem{CIP} 
  G.~P.~Holder, K.~M.~Nollett and A.~van Engelen,
  %``On Possible Variation in the Cosmological Baryon Fraction,''
  Astrophys.\ J.\  {\bf 716}, 907 (2010)
  [arXiv:0907.3919 [astro-ph.CO]];
  %%CITATION = ARXIV:0907.3919;%%
  %14 citations counted in INSPIRE as of 25 Mar 2013
  C.~Gordon and J.~R.~Pritchard
  %``Forecasted 21 cm constraints on compensated isocurvature perturbations,''
  Phys.\ Rev.\ D {\bf 80}, 063535 (2009)
  [arXiv:0907.5400 [astro-ph.CO]];
  %%CITATION = ARXIV:0907.5400;%%
  %11 citations counted in INSPIRE as of 25 Mar 2013
  D.~Grin, O.~Dore and M.~Kamionkowski,
  %``Do baryons trace dark matter in the early universe?,''
  Phys.\ Rev.\ Lett.\  {\bf 107}, 261301 (2011)
  [arXiv:1107.1716 [astro-ph.CO]].
  %%CITATION = ARXIV:1107.1716;%%
  %6 citations counted in INSPIRE as of 25 Mar 2013
  D.~Grin, O.~Dore and M.~Kamionkowski,
  %``Compensated Isocurvature Perturbations and the Cosmic Microwave Background,''
  Phys.\ Rev.\ D {\bf 84}, 123003 (2011)
  [arXiv:1107.5047 [astro-ph.CO]].
  %%CITATION = ARXIV:1107.5047;%%
  %2 citations counted in INSPIRE as of 25 Mar 2013


%\cite{Ade:2013nta}
\bibitem{Ade:2013nta} 
  P.~A.~R.~Ade {\it et al.}  (Planck Collaboration),
  %``Planck 2013 results. XVIII. Gravitational lensing-infrared background correlation,''
  arXiv:1303.5078 [astro-ph.CO];
  %%CITATION = ARXIV:1303.5078;%%
  P.~A.~R.~Ade {\it et al.}  (Planck Collaboration),
  %``Planck 2013 results. XVII. Gravitational lensing by large-scale structure,''
  arXiv:1303.5077 [astro-ph.CO];
  %%CITATION = ARXIV:1303.5077;%%
  (Planck Collaboration),B
  %``Planck 2013 results. XXI. Cosmology with the all-sky \Planck\ Compton parameter $y$-map,''
  arXiv:1303.5081 [astro-ph.CO].
  %%CITATION = ARXIV:1303.5081;%%

%\cite{Gibelyou:2012ri}
\bibitem{Gibelyou:2012ri} 
  C.~Gibelyou and D.~Huterer,
  %``Dipoles in the Sky,''
  Mon.\ Not.\ Roy.\ Astron.\ Soc.\  {\bf 427}, 1994 (2012)
  [arXiv:1205.6476 [astro-ph.CO]].
  %%CITATION = ARXIV:1205.6476;%%

\bibitem{Loeb:2003ya}
 A.~Loeb and M.~Zaldarriaga,
 %``Measuring the small - scale power spectrum of cosmic density fluctuations
 %through 21 cm tomography prior to the epoch of structure formation,''
 Phys.\ Rev.\ Lett.\  {\bf 92}, 211301 (2004)
 [arXiv:astro-ph/0312134];
 %%CITATION = PRLTA,92,211301;%%
  S.~Jester and H.~Falcke,
  %``Science with a lunar low-frequency array: from the dark ages of the Universe to nearby exoplanets,''
  New Astron.\ Rev.\ \ {\bf 53}, 1  (2009)
  [arXiv:0902.0493 [astro-ph.CO]].
  %%CITATION = ASTRE,53,1;%%

\bibitem{21cm}
 S.~Furlanetto, S.~P.~Oh and F.~Briggs,
 %``Cosmology at Low Frequencies: The 21 cm Transition and the High-Redshift
 %Universe,''
 Phys.\ Rept.\  {\bf 433}, 181 (2006)
 [arXiv:astro-ph/0608032];
 %%CITATION = PRPLC,433,181;%%
  M.~F.~Morales and J.~S.~B.~Wyithe,
  %``Reionization and Cosmology with 21 cm Fluctuations,''
  Annu.\ Rev.\ Astro.\ Astrophys.\ {\bf 48}, 127 (2010)
  [arXiv:0910.3010 [astro-ph.CO]].
  %%CITATION = ARXIV:0910.3010;%%

%\cite{Chluba:2012we}
\bibitem{Chluba:2012we} 
  J.~Chluba, R.~Khatri, and R.~A.~Sunyaev, 
  %``CMB at 2 ? 2 order: the dissipation of primordial acoustic waves and the observable part of the associated energy release,"
  MNRAS\ {\bf 425}, 1129 (2012);
  [arXiv:1202.0057 [astro-ph.CO]].
  %
  E.~Pajer and M.~Zaldarriaga,
  %``A New Window on Primordial non-Gaussianity,''
  Phys.\ Rev.\ Lett.\  {\bf 109}, 021302 (2012)
  [arXiv:1201.5375 [astro-ph.CO]].
  %%CITATION = ARXIV:1201.5375;%%
  %16 citations counted in INSPIRE as of 25 Mar 2013
  J.~Chluba, A.~L.~Erickcek and I.~Ben-Dayan,
  %``Probing the inflaton: Small-scale power spectrum constraints from measurements of the CMB energy spectrum,''
  Astrophys.\ J.\  {\bf 758}, 76 (2012)
  [arXiv:1203.2681 [astro-ph.CO]];
  %%CITATION = ARXIV:1203.2681;%%
  %13 citations counted in INSPIRE as of 25 Mar 2013


%\cite{Kamionkowski:1996zd}
\bibitem{Kamionkowski:1996zd} 
  M.~Kamionkowski, A.~Kosowsky and A.~Stebbins,
  %``A Probe of primordial gravity waves and vorticity,''
  Phys.\ Rev.\ Lett.\  {\bf 78}, 2058 (1997)
  [astro-ph/9609132];
  %%CITATION = ASTRO-PH/9609132;%%
  %311 citations counted in INSPIRE as of 25 Mar 2013
  U.~Seljak and M.~Zaldarriaga,
  %``Signature of gravity waves in polarization of the microwave background,''
  Phys.\ Rev.\ Lett.\  {\bf 78}, 2054 (1997)
  [astro-ph/9609169].
  %%CITATION = ASTRO-PH/9609169;%%
  %342 citations counted in INSPIRE as of 25 Mar 2013

%\cite{Hajian:2004zn}
\bibitem{Hajian:2004zn} 
  A.~Hajian, T.~Souradeep and N.~J.~Cornish,
  %``Statistical isotropy of the WMAP data: A Bipolar power spectrum analysis,''
  Astrophys.\ J.\  {\bf 618}, L63 (2004)
  [astro-ph/0406354].
  %%CITATION = ASTRO-PH/0406354;%%
  %66 citations counted in INSPIRE as of 25 Mar 2013

%\cite{Jeong:2012df}
\bibitem{Jeong:2012df} 
  D.~Jeong and M.~Kamionkowski,
  %``Clustering Fossils from the Early Universe,''
  Phys.\ Rev.\ Lett.\  {\bf 108}, 251301 (2012)
  [arXiv:1203.0302 [astro-ph.CO]].
  %%CITATION = ARXIV:1203.0302;%%
  %6 citations counted in INSPIRE as of 22 Mar 2013

%\cite{Book:2011na}
\bibitem{Book:2011na} 
  L.~G.~Book, M.~Kamionkowski and T.~Souradeep,
  %``Odd-Parity Bipolar Spherical Harmonics,''
  Phys.\ Rev.\ D {\bf 85}, 023010 (2012)
  [arXiv:1109.2910 [astro-ph.CO]].
  %%CITATION = ARXIV:1109.2910;%%
  %10 citations counted in INSPIRE as of 25 Mar 2013

%\cite{Dai:2013ikl}
\bibitem{Dai:2013ikl} 
  L.~Dai, D.~Jeong and M.~Kamionkowski,
  %``Seeking Inflation Fossils in the Cosmic Microwave Background,''
  arXiv:1302.1868 [astro-ph.CO].
  %%CITATION = ARXIV:1302.1868;%%


%\cite{Chang:2008gj}
\bibitem{Chang:2008gj} 
  S.~Chang, M.~Kleban and T.~S.~Levi,
  %``Watching Worlds Collide: Effects on the CMB from Cosmological Bubble Collisions,''
  JCAP {\bf 0904}, 025 (2009)
  [arXiv:0810.5128 [hep-th]].
  %%CITATION = ARXIV:0810.5128;%%
  %37 citations counted in INSPIRE as of 20 Mar 2013

%\cite{Ade:2013vbw}
\bibitem{Ade:2013vbw} 
  P.~A.~R.~Ade {\it et al.}  [ Planck Collaboration],
  %``Planck 2013 results. XXVI. Background geometry and topology of the Universe,''
  arXiv:1303.5086 [astro-ph.CO].
  %%CITATION = ARXIV:1303.5086;%%

\end{thebibliography}
\end{document}